\documentstyle[prd,aps,epsfig,psfig]{revtex}
\tighten
\begin{document}
\draft
\title{Cosmological scaling solutions of non-minimally coupled scalar 
fields}
\author{Jean-Philippe Uzan}
\address{D\'epartement de Physique Th\'eorique, Universit\'e
de Gen\`eve,\\
24 quai E. Ansermet, CH-1211 Geneva 4 (Switzerland).\\
and\\
D\'epartement d'Astrophysique Relativiste et de Cosmologie,\\
Observatoire de Paris-Meudon, UPR 176, CNRS, 92195 Meudon (France).}
\date{\today}
\maketitle
\begin{abstract}
We study the existence and stability of cosmological scaling solutions
of a non-minimally coupled scalar field evolving in either an
exponential or inverse power law potential.  We show that for inverse
power law potentials there exist scaling solutions the stability of
which does not depend on the coupling constant $\xi$. We then study
the more involved case of exponential potentials and show that the
scalar field will assymptotically behaves as a barotropic fluid when
$\xi\ll1$. The general case $\xi\not\ll1$ is then discussed an we
illustrate these results by some numerical examples.
\end{abstract}
\pacs{PACS numbers: 98.80.Cq\\
Preprint number: UGVA-DPT 1998/12-1022}
\vskip2pc

\section{Introduction}\label{I}

 On one hand recent cosmological observations, and
particuarly the Hubble diagram for Type Ia supernovae \cite{perlmutter97}
have led to the idea that the universe may be dominated by a
component with negative pressure \cite{zlatev98} and thus that today the
universe is accelerating. Such a component can also, if one
sticks to the prediction of inflation that $\Omega=1$, account
for the ``missing energy''. Yet, many candidates have been 
proposed like a cosmological constant, a ``dynamical''
cosmological constant \cite{coble96}, cosmic strings \cite{spergel96}
or a spatially homogeneous scalar field rolling down a potential
\cite{huey98}.

On the other hand, potentials decreasing to zero for infinite value of
the field have been shown to appear in particle physics models (see
e.g. \cite{binetruy97,binetruy98}). For instance, exponential
potentials arise in high order gravity \cite{barrow88}, in
Kaluza-Klein theories which are compactified to produce the four
dimensional observed universe \cite{wetterich85} or can arise due to
non-perturbative effects such as gaugino condensation
\cite{carlos93}. Inverse power law potentials can be obtained in
models where supersymmetry is broken through fermion condensates
\cite{binetruy98}.  This gives one more theoretical motivation to study
the cosmological implications of a field with such potentials.

The cosmological solutions with such a field have first been studied
by Ratra and Peebles \cite{ratra88} (see also
\cite{wands93,wetterich94}) who showed the existence and stability of
scaling solutions in respectively a field, radiation or matter
dominated dominated universe for a field evolving in an exponential
and inverse power law potential.  A complete study in the framework of
barotropic cosmologies in the case of the exponential potential
\cite{copeland97,billyard98} show that the solutions were stable to
shear perturbations and to curvature perturbations when $P/\rho<-1/3$,
but that for realistic matter (such as dust) these solutions were
unstable, essentialy to curvature perturbations. Liddle and Scherrer
\cite{liddle98} made a complete classification of the field potentials
and show that power law potentials also lead to scaling solutions
(i.e. to solutions such that the field energy density $\rho_\phi$,
behaves as the scale factor at a given power) and the coupling of the
field to ordinary matter has been considered in \cite{carroll98}. Such
solutions are indeed of interest in cosmology since they provide a
candidate for a component with negative pressure.

Cosmological models with a scalar field have started to be investigated
\cite{perrotta98,peebles98} for different kind of potentials as cosine
potential \cite{coble96}, exponential potential
\cite{ferreira97,viana98} and inverse power law potentials
\cite{caldwell97}. It has also been shown
that the luminosity distance as a function of redshift
\cite{huterer98,starobinski98} or the behaviour of density
perturbations in the dust era as a function of redshift
\cite{starobinski98} can be used to reconstruct the field potential.

However, all these studies have been done under the hypothesis that
the field is minimally coupled to the metric. It is known that terms
with such a non-minimal coupling $\bar{\cal R}f(\phi)$ between the
scalar curvature $\bar{\cal R}$ and the field $\phi$ can appear when
quantizing fields in curved spacetime \cite{birrel82,ford87}, in
multi-dimensional theories \cite{maeda86} like superstring and induced
gravity theories \cite{acetta85}.  Since these theories predict both
the existence of scalar fields with potential or power-law potential
and non-minimal coupling, it is of interest to study the influence of
this coupling, and for instance the robustness of the existence and
stability of scaling solutions.  The influence of such a coupling
during an inflationary period and the existence of inflationnary
attractors have yet been examined (see e.g. \cite{amendola}).

In this article, we study the stability of scaling solutions of a
non-minimally coupled scalar field. We first present (\S \ref{II}) the
main notations and equations.  After having, in \S \ref{III}, briefly
recalled the standard approach for determining the potentials that can
give rise to such behaviour for a minimally coupled scalar field, we
investigate the inverse power law potentials (\S \ref{IV}) and the
exponential potential (\S V). In the latter case, we study the two limiting
situations $\xi\ll1$ and $\xi\gg1$ and then have a heuristic discussion
in the general case.

\section{Description of the model}\label{II}

We assume that the universe is described by a Friedmann-Lema\^\i tre
model with Euclidean spatial sections so that the line element reads
\begin{equation}
ds^2=-dt^2+a^2(t)\left[\delta_{ij}dx^idx^j\right]
\equiv {\bar g}_{\mu\nu}dx^\mu dx^\nu,\label{metrique_bgd}
\end{equation}
where $a(t)$ is the scale factor and  $t$ the cosmic time.
Greek indices are running from 0 to 3 and latin indices from 1 to 3.
The Hubble parameter H is defined as $H\equiv \dot a/a$,
where a dot denotes a derivative with respect to $t$. We also
introduce $\bar\nabla_\mu$ the covariant derivative associated to
${\bar g}_{\mu\nu}$.

We assume that the matter content of this universe is composed of a
perfect fluid and a homogeneous scalar field $\phi$ coupled to gravity
and coupled to matter only through gravity. The fluid energy density,
$\rho_{_B}$, and pressure, $P_{_B}$, are related through the
equation of state
\begin{equation}
P_{_B}=\omega_{_B}\rho_{_B},
\end{equation}
where $B$ refers to ``background''. The conservation of the
energy-momentum of the fluid reduces to
\begin{equation}
\dot\rho_{_B}+3H(\rho_{_B}+P_{_B})=0.\label{cons_fluide}
\end{equation}

The scalar field $\phi$ evolves in a potential $V(\phi)$ and its
dynamics is given by the Lagragian
\begin{equation}
S_\phi=-\frac{1}{2}\int\left[\partial_\mu\phi\partial^\mu\phi
+2\xi \bar{\cal R}f(\phi)+2V(\phi)\right]\sqrt{-\bar g}d^4x,
\label{action}
\end{equation}
where $\xi$ is the field-metric coupling constant ($\xi=0$ corresponds
to a minimally coupled field and $\xi=1/6$ to a conformally coupled
field), $\bar{\cal R}$ is the scalar curvature of the spacetime.  No
known fundamental principle predicts the functional form $f(\phi)$ and
we will assume that $f(\phi)=\phi^2/2$. This is however the only choice that
allows for a dimensionless $\xi$. The equation of evolution is
then obtained by varying the action (\ref{action}) with respect to the
field which leads to the Klein-Gordon equation
\begin{equation}
\frac{\delta S_\phi}{\delta\phi}=0\Longleftrightarrow \Box\phi-\xi
\bar{\cal R}\phi-a^2\frac{dV}{d\phi}=0,\label{evolution1}
\end{equation}
where $\Box\equiv\bar\nabla_\mu\bar\nabla^\mu$. For an homogeneous
field in the spacetime (\ref{metrique_bgd}), it reduces to
\begin{equation}
\ddot\phi+3H\dot\phi +\frac{dV}{d\phi} +6\xi\left(2H^2+\dot H\right)
\phi=0.\label{klein_gordon}
\end{equation}
The energy density and the pressure of this field are defined by
\begin{equation}
\rho_\phi\equiv a^{-2}T_{00};\qquad P_\phi\equiv \frac{1}{3}
a^{-2}T_{ij}\delta^{ij}\label{defPrho}
\end{equation}
and we define $\omega_\phi\equiv P_\phi/\rho_\phi$.  The
energy-momentum tensor $T_{\mu\nu}$ is obtained by varying the action
(\ref{action}) with respect to the metric $\bar g_{\mu\nu}$ and reads
\begin{eqnarray}
T_{\mu\nu}&=& (1-2\xi)\bar\nabla_\mu\phi\bar\nabla_\nu\phi +
\left(2\xi-\frac{1}{2}\right){\bar g}_{\mu\nu}\bar\nabla_\lambda\phi\bar
\nabla^\lambda\phi
-2\xi\phi \bar\nabla_\mu\bar\nabla_\nu\phi
+2\xi\phi\Box\phi{\bar g}_{\mu\nu}+\xi
{\bar G}_{\mu\nu}\phi^2 - V(\phi){\bar g}_{\mu\nu},
\end{eqnarray}
$G_{\mu\nu}$ being the Einstein tensor of the background metric.  This
expression can be compared to standard results (see e.g.
\cite{birrel82}). Note however that when the signature of the metric is
$(+,-,-,-)$, one has to change the sign of $\bar g_{\mu\nu}$, $\Box$, $\bar
R_{\mu\nu}$, and $\bar G_{\mu\nu}$ while $T_{\mu\nu}$ and $\bar{\cal
R}$ remain unaffected.

It is then straigthforward to check that the density and the
pressure (\ref{defPrho}) of the scalar field are given by
\begin{eqnarray}
\rho_\phi&=&\frac{1}{2}\dot\phi^{2}+V(\phi)+3H\xi
\phi(2\dot\phi+ H\phi)\label{rhophi}\\
P_\phi&=&\frac{1}{2}\dot\phi^{2}-V(\phi)-\xi\left((2{\dot H}+3{H}^2)\phi^2
+4{H}\phi\dot\phi+2\phi\ddot\phi+2\dot\phi^{2}\right),
\label{pphi}
\end{eqnarray}
and that the conservation of the energy-momentum of the field
($\bar\nabla_\mu T^{\mu\nu}=0$) reduces to the Klein-Gordon equation
(\ref{klein_gordon})
\begin{equation}
\dot\rho_\phi+3{H}(\rho_\phi+P_\phi)=0\Longleftrightarrow
\ddot\phi+3{H}\dot\phi +\frac{dV}{d\phi} +6\xi\left(2{H}^2+{\dot H}
\right)\phi=0.\label{phiflu}
\end{equation}
[We have used that the scalar curvature and the Einstein tensor 
are respectively given by $\bar{\cal R}=6\left(2H^2+\dot H\right)$,
$\bar G_{00}=3a^2H^2$ and $\bar G_{ij}=-(2\dot H+3H^2)a^2\delta_{ij}$].
The equation of state of the field is defined by
\begin{equation}
\omega_\phi=\frac{P_\phi}{\rho_\phi}.
\end{equation}

The matter content being described, we can write the Einstein
equations which dictate the dynamics of the spacetime and, in our
case, reduces to the Friedmann equations
\begin{eqnarray}
{H}^2&=&\frac{\kappa}{3}\left(\rho_{_B}+\rho_\phi\right),\label{friedmann}\\
\dot H &=& -\frac{\kappa}{2}\left((\omega_{_B}+1)\rho_{_B}+\rho_\phi+P_\phi
\right)\label{friedmann2}
\end{eqnarray}
with $\kappa=8\pi G$. One equation of the set
(\ref{cons_fluide},\ref{phiflu}, \ref{friedmann},\ref{friedmann2}) is
redundant due to the Bianchi identities. It will also be usefull to
introduce the density parameter of a component $X$ as $\Omega_X\equiv
\kappa\rho_X/3H^2$.  They must satisfy (from equation \ref{friedmann})
the constraint
\begin{equation}
\Omega_{_B}+\Omega_\phi=1.
\end{equation}

\section{Scaling solutions for minimally coupled scalar fields ($\xi=0$)}\label{III}

In this section, we briefly recall the ``standard'' procedure to show
that there exist scaling solutions and to determine the potentials
that give rise to such solutions. This presentation will also enable
us to understand the differences with the more general case of
non-minimally coupled scalar fields. We follow the approach initiated
by Ratra and Peebles \cite{ratra88} and others
\cite{wetterich94,liddle98} where one, assuming a scaling form for
$\rho_\phi$, derives an equation for $\phi(a)$ and then uses
(\ref{rhophi}-\ref{pphi}) to deduce the associated potential.

The equation (\ref{cons_fluide}) of evolution of the background fluid
density can be integrated to give
\begin{equation}
\rho_{_B}=\rho_{_{B0}} x^{-m}\quad\hbox{with}\quad x\equiv \frac{a}{a_0},
\end{equation}
where a subscript 0 refers to quantities evaluated at a given
initial time. We now look for scaling solutions, i.e. solutions such
that
\begin{equation}
\rho_\phi=\rho_{\phi0} x^{-n}\Longleftrightarrow
 P_\phi=\left(\frac{n}{3}-1\right)\rho_\phi.
\end{equation}
Since $n/3-1=1-2V/\rho_\phi\in[-1;1]$, we deduce that $n\in[0,6]$ (let
us emphasize that this is a priori no longer true when $\xi\not=0$).
Using equations (\ref{rhophi}-\ref{pphi}), such a solution must
satisfy
\begin{equation}
\dot\phi^2=\frac{n}{3}\rho_\phi\quad\hbox{and}\quad V(\phi)=
\left(1-\frac{n}{6}\right)\rho_\phi.\label{phistate}
\end{equation}
Now, the Friedmann equation (\ref{friedmann}) implies that the field
should satisfy
\begin{equation}
\frac{d\phi}{dx}=\frac{A}{x\sqrt{1+B^2x^{n-m}}}\quad\hbox{with}\quad
B\equiv\sqrt{\frac{\rho_{B0}}{\rho_{\phi0}}}\quad\hbox{and}\quad
A\equiv\sqrt{\frac{n}{\kappa}}\sqrt{\frac{\rho_{\phi0}}
{\rho_{\phi0}+\rho_0}}=\sqrt{\frac{n}{\kappa}\Omega_{\phi0}}.
\label{dphidx}
\end{equation}
This can be integrated for different relative values of $m$ and $n$.

\subsection{$m=n$ case}\label{exp} 

In that situation, (\ref{dphidx}) leads to
\begin{equation}
\phi-\phi_0=\frac{n}{\lambda} \ln{x}\quad\hbox{with}\quad
\lambda^{-1}\equiv\frac{\Omega_{\phi0}}{\sqrt{n\kappa}}
\end{equation}
and then using (\ref{phistate}), to the potential
\begin{equation}
V(\phi)=\left(1-\frac{n}{6}\right)\rho_{\phi0}\hbox{e}^{
-\lambda(\phi-\phi_0)}.
\end{equation}

This solution corresponds to the scalar field dominated universe of
Ratra and Peebles \cite{ratra88} and to their scaling solution in the
case $m=3$ and $m=4$ (i.e. radiation or matter dominated
universe). Note that with such a potential, $\rho_\phi$ will, by
construction, mimic the evolution of the background fluid and that we
do not have to assume that $\Omega_\phi\ll1$.

Note however that, if the scalar field has reached the attractor from
very early time, $\rho_\phi$ behaves like radiation and thus
contributes to a non-negligible part of the radiation content during
the nucleosynthesis and it has been shown that it implies the
constraint $\Omega_{\phi0}<0.15$ \cite{ferreira97,wands93}. Moreover,
since $\omega_\phi=0$ in the matter era, such a field will not explain
the supernovae measurements (which seem to favor $\omega_\phi=-0.6$
\cite{turner97}) even if it can account for a substantial part of the
dark matter.

\subsection{$m\not=n$ case}\label{IIIplaw}

In this case we have
\begin{equation}
\frac{d\phi}{dx}=\frac{A}{x\sqrt{1+B^2x^{n-m}}},
\end{equation}
which can be integrated ($B>0$) to give \cite{gradshteyn}
\begin{equation}
\phi-\phi_0=\frac{2A}{m-n}\ln{\left[
\sqrt{1+\left(B^{-1}x^{\frac{m-n}{2}}\right)^2}+B^{-1}x^{\frac{m-n}{2}}
\right]}.\label{xpot}
\end{equation}
Again, using (\ref{phistate}), we can deduce the potential
\begin{equation}
V(\phi)=\left(1-\frac{n}{6}\right)\rho_{\phi0}x^{-n},\label{xpot2}
\end{equation}
$x$ being given by (\ref{xpot}). Indeed, we only get the potential
in a parametric form, but when $B\gg1$ (i.e. when the perfect fluid
drives the evolution of the universe, $x$ can be eliminated from 
(\ref{xpot}-\ref{xpot2}) to give
\begin{equation}
V(\phi)=\left(1-\frac{n}{6}\right)\rho_{\phi0}
\left(\frac{m-n}{2A}B\right)^{-\frac{2n}{m-n}}
(\phi-\phi_0)^{-\frac{2n}{m-n}}
\end{equation}
When $m=3$ and $m=4$, we recover the Ratra-Peebles result
\cite{ratra88} as well as the Liddle-Scherrer result \cite{liddle98}
for all $m$.  This parametric general form of the potential seems
however not to have been exhibited before.

\section{Non-minimally coupled scalar fields with a power
law potential}\label{IV}

\subsection{Existence of a scaling solution}

The former procedure cannot be applied when the field is non-minimally
coupled since it is impossible for instance to write a closed equation
for $\frac{d\phi}{dx}$ as in (\ref{dphidx}).  Moreover, we are
interested in a field evolving in a given potential.  We assume that
the field evolves in an inverse power law potential and show that
there exist scaling solutions, the stability of which is then studied.

We assume that the potential takes the form
\begin{equation}
V(\phi)=V_0M_p^4\left(\frac{\phi}{M_p}\right)^{-\alpha}\quad\hbox{with}\quad \alpha>0,
\end{equation}
with $M_p$ being the Planck mass.
The universe is dominated by the perfect fluid so that (we assume $m\not=0$)
\begin{equation}
H=\frac{2}{m}\frac{1}{t-t_0};\quad a=a_0(t-t_0)^{2/m};\quad
\omega_{_B}=\frac{m}{3}-1.
\end{equation}
Redefining $M_p(t-t_0)\sqrt{V_0}$ as $t$ and $\phi/M_p$ as $\phi$, the
Klein-Gordon equation (\ref{klein_gordon}) becomes
\begin{equation}
\ddot\phi+\frac{6}{m}\frac{1}{t}\dot\phi+\frac{12}{m}\left(\frac{4}{m}-1\right)\xi
\frac{1}{t^2}\phi-\alpha\phi^{-(\alpha+1)}=0.\label{kg_law}
\end{equation}
Looking for a solution of this equations of the form $\phi\propto t^\beta$,
one obtains 
\begin{equation}
\phi=\phi_0t^\beta,\quad \phi_0^{\alpha+2}=\frac{\alpha}{\beta
\left(\beta+\frac{6}{m}-1\right)+\frac{12}{m}\left(\frac{4}{m}-1\right)\xi},
\quad \beta=\frac{2}{\alpha+2},\label{scaling_plaw}
\end{equation}
so that $\rho_\phi\propto a^{-n}$ with $n/m=\alpha/(\alpha+2)$.  This
solution is only well-defined if
\begin{equation}
\frac{6}{m}-\frac{\alpha}{\alpha+2}+\frac{6}{m}\left(\frac{4}{m}-1\right)
(\alpha+2) \xi>0.\label{cond}
\end{equation}

One can then compute the energy and the pressure of this field by
inserting this solution into (\ref{rhophi}-\ref{pphi}) and verify, after
some algebra, that
\begin{equation}
\omega_\phi=\frac{\omega_{_B}\alpha-2}{\alpha+2},
\end{equation}
whatever the value of $\xi$. This shows that the scaling solution does
not depend on the coupling in the sense that $\omega_\phi$ is
independent of $\xi$. This relation generalises the one found for
minimally-coupled scalar fields \cite{zlatev98,ratra88,liddle98}.

\subsection{stability}

As emphasized in the previous section, the scaling solution does not
depend on the coupling $\xi$. The stability of such a solution is known
when $\xi=0$ \cite{ratra88,liddle98},  we now have to study it when
$\xi\not=0$. Following \cite{ratra88}, we define the new set of variables
\begin{eqnarray}
t&=&\hbox{e}^\tau\\
u(\tau)&=&\frac{\phi(\tau)}{\phi_s(\tau)},
\end{eqnarray}
where $\phi_s$ is the scaling solution (\ref{scaling_plaw}).
We set $\phi'\equiv \frac{d\phi}{d\tau}$. Using
$\dot\phi=\hbox{e}^{-\tau}\phi'$, $\ddot\phi=\hbox{e}^{-2\tau}(\phi''-\phi')$ and
$\phi_s'/\phi_s=2/(\alpha+2)$,
equation (\ref{kg_law}) reduces to
\begin{equation}
u''+\left(\frac{6}{m}+\frac{4}{\alpha+2}-1\right)u'+
\left(\frac{2}{\alpha+2}\left[\frac{6}{m}-\frac{\alpha}{\alpha+2}\right]
+\frac{12}{m}\left[\frac{4}{m}-1\right]\xi\right)\left(u-u^{-(\alpha+1)}\right)=0.
\label{eq_lambda}
\end{equation}
The scaling solution corresponds to the critical point $u=1$. Introducing
$v=u'$ and linearising around this critical point [$u=1+\epsilon$], we obtain
\begin{equation}
\left(
\begin{array}{l}
\epsilon \\
v
\end{array}
\right)'=
\left(
\begin{array}{cc}
0&1\\
-2\left(\left[\frac{6}{m}-\frac{\alpha}{\alpha+2}\right]
+\frac{6(\alpha+2)}{m}\left[\frac{4}{m}-1\right]\xi\right)
&\left(1-\frac{6}{m}-\frac{4}{\alpha+2}\right)
\end{array}
\right)
\left(
\begin{array}{l}
\epsilon \\
v
\end{array}
\right).
\end{equation}
The eigenvalues, $\lambda_\pm$, of this system are 
\begin{equation}
2\lambda_\pm=\left(1-\frac{6}{m}-\frac{4}{\alpha+2}\right)\pm
\sqrt{\left(1-\frac{6}{m}-\frac{4}{\alpha+2}\right)^2
-8\left[\left(\frac{6}{m}-\frac{\alpha}{\alpha+2}\right)
+\frac{6}{m}(\alpha+2)\left(\frac{4}{m}-1\right)\xi\right]}
\label{racine_lambda}
\end{equation}
This expression reduces to the Liddle-Scherrer
result \cite{liddle98} when $\xi=0$ [with $\alpha/(\alpha+2)=n/m$] and to the
Ratra-Peebles result \cite{ratra88} when either $m=3$ or $m=4$.

A necessary and sufficient condition for the critical point to be
stable, is the negativity of the real part of the two eigenvalues.
Defining $\bar\xi$ as
\begin{equation}
\bar\xi=\frac{\left(m-6-\frac{4m}{\alpha+2}\right)^2-8m\left(6-\frac{\alpha m}{\alpha+2}
\right)}{48(\alpha+2)(4-m)},
\end{equation}
the two solutions of (\ref{racine_lambda}) are real when $(\xi-\bar\xi)(m-4)\geq0$. Thus,
\begin{itemize}
\item when the two eigenvalues are complex, $\lambda_=\lambda_+^*$ and their real part
is given by, $Re(\lambda_\pm)=\left(1-\frac{6}{m}-\frac{4}{\alpha+2}
\right)/2$ and the scaling solution is stable if
\begin{equation}
1-\frac{6}{m}-\frac{4}{\alpha+2}<0.
\end{equation}
\item when the two eigenvalues $\lambda_\pm$ are real, their
product is (from \ref{eq_lambda})
\begin{equation}
\lambda_-\lambda_+=2\left(\left[\frac{6}{m}-\frac{\alpha}{\alpha+2}\right]
+\frac{6(\alpha+2)}{m}\left[\frac{4}{m}-1\right]\xi\right)>0,
\end{equation}
because of the condition (\ref{cond}). They are thus of the same sign, their
sum being
\begin{equation}
\lambda_++\lambda_-=1-\frac{6}{m}-\frac{4}{\alpha+2},
\end{equation}
they will both be negative if $\lambda_++\lambda_-<0$ and thus,
the solution will be stable only if 
\begin{equation}
1-\frac{6}{m}-\frac{4}{\alpha+2}<0,
\end{equation}
as in the case $\xi\geq\bar\xi$.
\end{itemize}

In conclusion, we find that {\it whatever} the coupling constant 
$\xi$, the scaling solution (\ref{scaling_plaw}) will be stable if and only if
\begin{equation}
1-\frac{6}{m}-\frac{4}{\alpha+2}<0.
\end{equation}
The value of $\xi$ only determines the nature of the stable point,
i.e. wether it is a stable spiral or a stable node.  This
generalises the study by Liddle-Scherrer \cite{liddle98} to a
non-minimally coupled scalar field.

\section{Non-minimally coupled scalar field in an exponential 
potential}\label{V}

\subsection{Scaling solutions ?}

\begin{figure}
\centering
\epsfig{figure=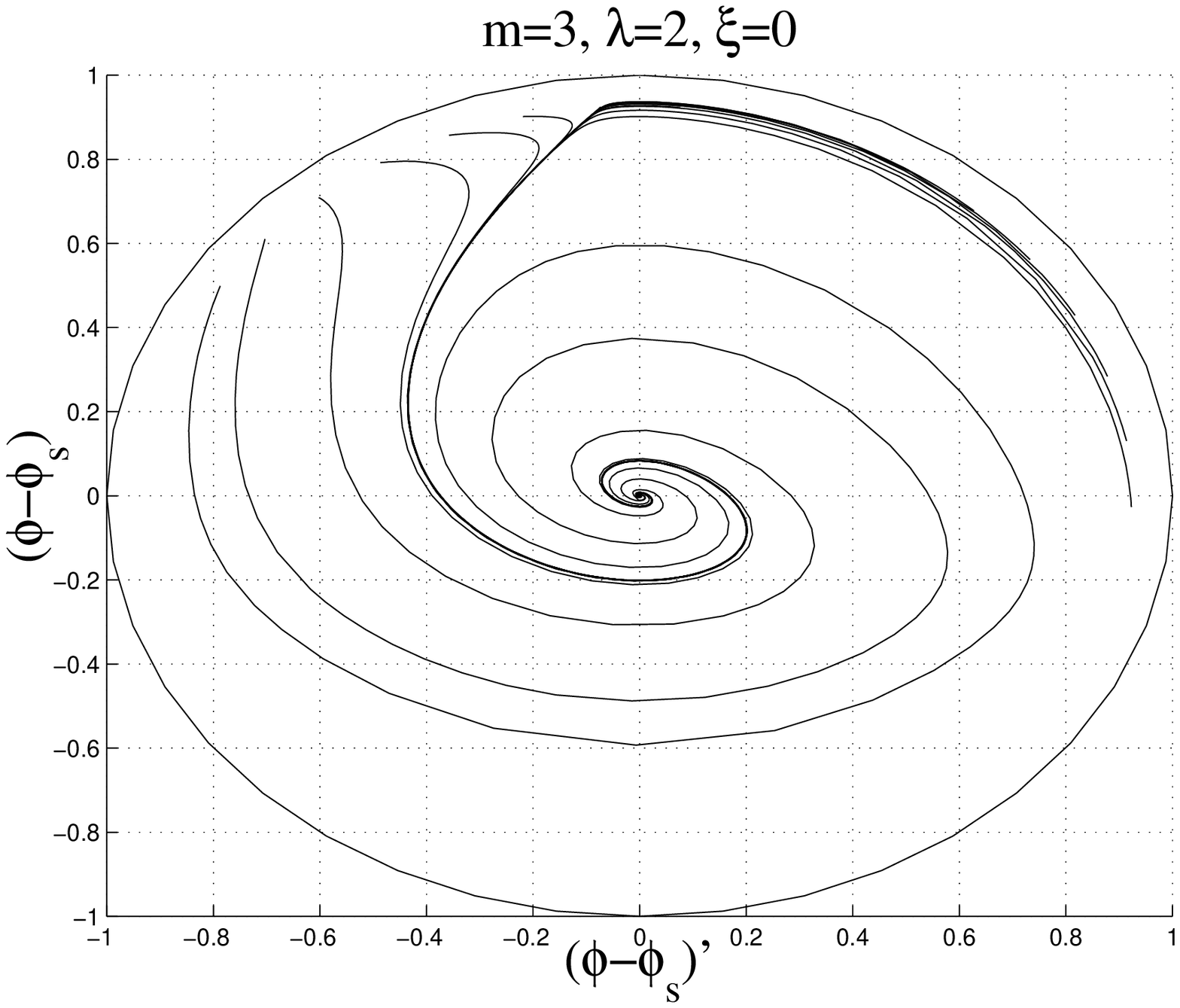, width=6cm}
\epsfig{figure=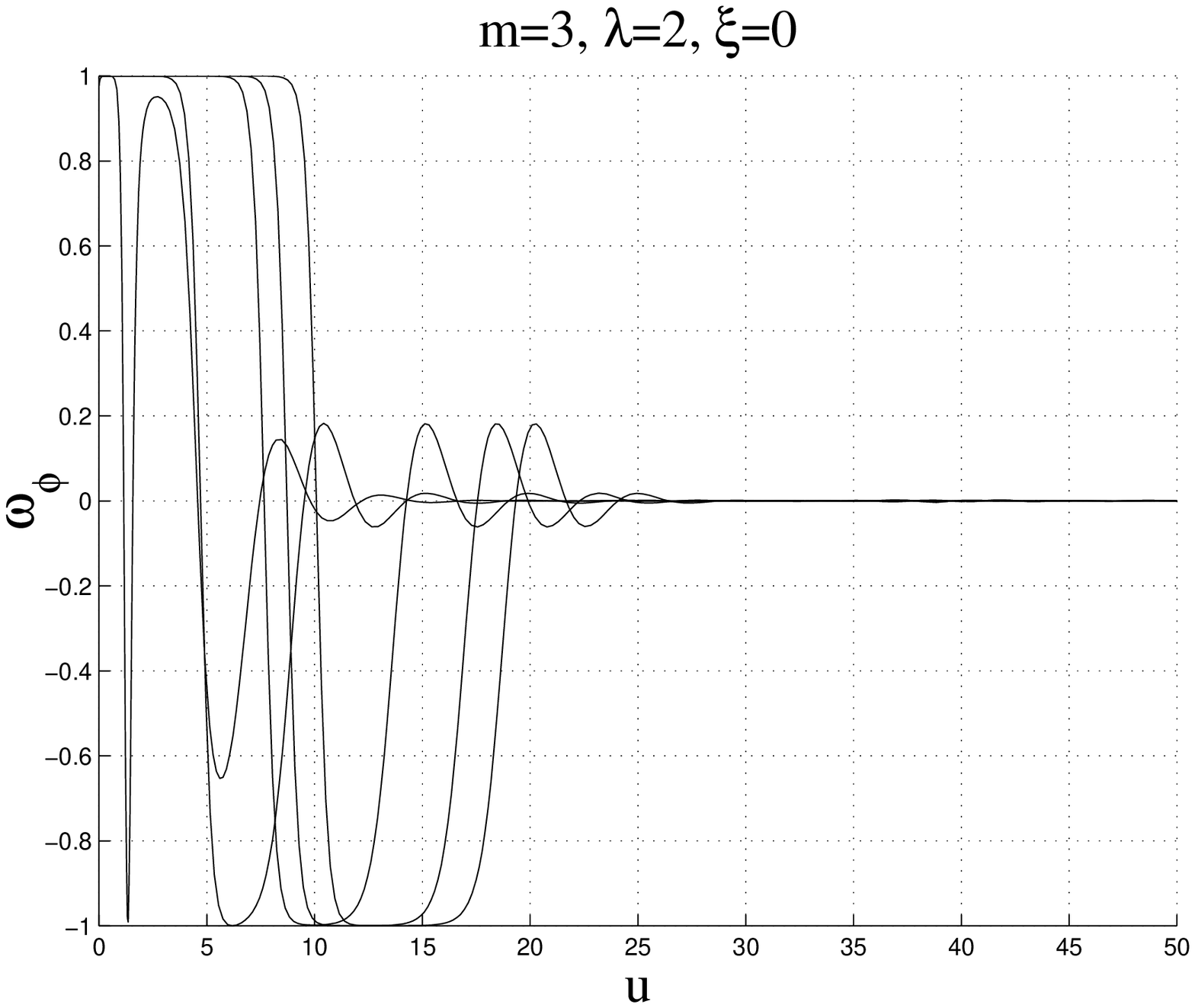, width=6cm}
\caption{ (left) The convergence towards the scaling solution 
when $\xi=0$. (right) The convergence towards the equation of
state $\omega_\phi=0$ for different initial conditions.}
\label{fig1}
\end{figure}

We now focuse on potentials of the form
\begin{equation}
V(\phi)=V_0M_p^4\hbox{e}^{-\lambda\phi/M_p}\quad\hbox{with}\quad
\lambda>0
\end{equation}
and work under the same assumptions as in \S \ref{IV}. Redefining $t$ and
$\phi$ as in
\S \ref{IV}, the Klein-Gordon equation now reads
\begin{equation}
\ddot\phi+\frac{6}{m}\frac{1}{t}\dot\phi+\frac{12}{m}\left(\frac{4}{m}-1\right)\xi
\frac{1}{t^2}\phi-\lambda\hbox{e}^{-\lambda\phi}=0.\label{kg_exp}
\end{equation}
When the coeficient of $\phi$ vanishes, i.e.  in a radiation dominated
universe ($m=4$) or when $\xi=0$, one can find a special solution of
the form $\phi=\ln{(At^\beta)}$. We get that (see \S \ref{exp}
with $\kappa=M_p^{-2}$)
\begin{equation}
\phi_s=\ln{(At^\beta)};\quad
A^{-\lambda}=\frac{2}{\lambda^2}\left(\frac{6}{m}-1\right) ;\quad
\beta=\frac{2}{\lambda}\Rightarrow \omega_{\phi_s}=\frac{m}{3}-1,
\label{sol_exp}
\end{equation}
where the subscript $s$ refers to the scaling solution.  In radiation
era, this implies for instance that $\omega_{\phi_s}=1/3$ and the
scalar field behaves like radiation.  Indeed, this is a very special
case since in that period $\bar{\cal R}=0$ and the field does not
``feel'' the non-minimal coupling and it evolves as if it were minimally
coupled. The complete study of these solution in function of the two
parameters ($\lambda,m$) \cite{copeland97,billyard98} shows that
when $\lambda^2>m$, the scaling solution $\phi_s$ is a stable node or
spiral whereas, when $\lambda^2<m$, the late time attractor is the
field dominated solution, which we do not consider here. The
convergence towards the solution $\phi_s$ is illustrated on figure
\ref{fig1}.

Now, in the most general case where $m\not=4$, it is easy to realize
that a solution of the form $\phi=\ln{(At^\beta)}$ cannot be solution
of equation (\ref{kg_exp}).

\subsubsection{The $|\xi|\ll1$ case}

Let us first look at the effect of a small perturbation in $\xi$ in
the sense that the potential term dominates over the coupling term in
the Klein-Gordon equation. For that purpose, we set
\begin{eqnarray}
u&\equiv&\ln{t}\nonumber\\
\phi&=&\phi_s+\xi\psi+{\cal O}(\xi^2).
\end{eqnarray}
The equation of evolution for $\psi$ can be deduced
from the Klein-Gordon equation (\ref{kg_exp})
\begin{equation}
\psi''+\left(\frac{6}{m}-1\right)\psi'+\frac{12}{m}
\left(\frac{4}{m}-1\right)\xi\psi
=-\frac{\lambda}{\xi}\left[1-\hbox{e}^{-\lambda\xi\psi}
\right]\hbox{e}^{2u-\lambda\phi_s}-\frac{12}{m}
\left(\frac{4}{m}-1\right)\phi_s,
\label{exp_int}
\end{equation}
where a prime denotes a derivative with respect to $u$.
Now, if we restrict to $\xi\ll1$ and linearise this equation
using the expression (\ref{sol_exp}) for $\phi_s$, we obtain at
zeroth order in $\xi$
\begin{equation}
\psi''+\left(\frac{6}{m}-1\right)\psi' +2\left(\frac{6}{m}-1\right)\psi=
\frac{12}{m\lambda}\left(\frac{4}{m}-1\right)\left[\ln{A^{-\lambda}}
-2u\right],
\end{equation}
the solution of which has the general form
\begin{equation}
\psi=B_+\hbox{e}^{\alpha_+u}+B_-\hbox{e}^{\alpha_-u}
-\frac{12}{m\lambda}\frac{4-m}{6-m}\left[u-\frac{1+\ln{A^{-\lambda}}}{2}
\right].\label{toto}
\end{equation}
$B_+\hbox{e}^{\alpha_+u}$ and $B_-\hbox{e}^{\alpha_-u}$ are two
independent solutions of the homogeneous equation.  Since
$\alpha_+\alpha_-=2(6/m-1)=-2(\alpha_++\alpha_-)$, we deduce that if
$m<6$, the real parts of both $\alpha_+$ and $\alpha_-$ are negative,
so that the two homogeneous solutions correspond to decaying modes and
the particular solution is then an attractor (if $m<2/3$ then
$\alpha_+$ and $\alpha_-$ are real, otherwise they are complex and the
homogeneous part will decay while oscillating).\\

\begin{figure}
\centering
\epsfig{figure=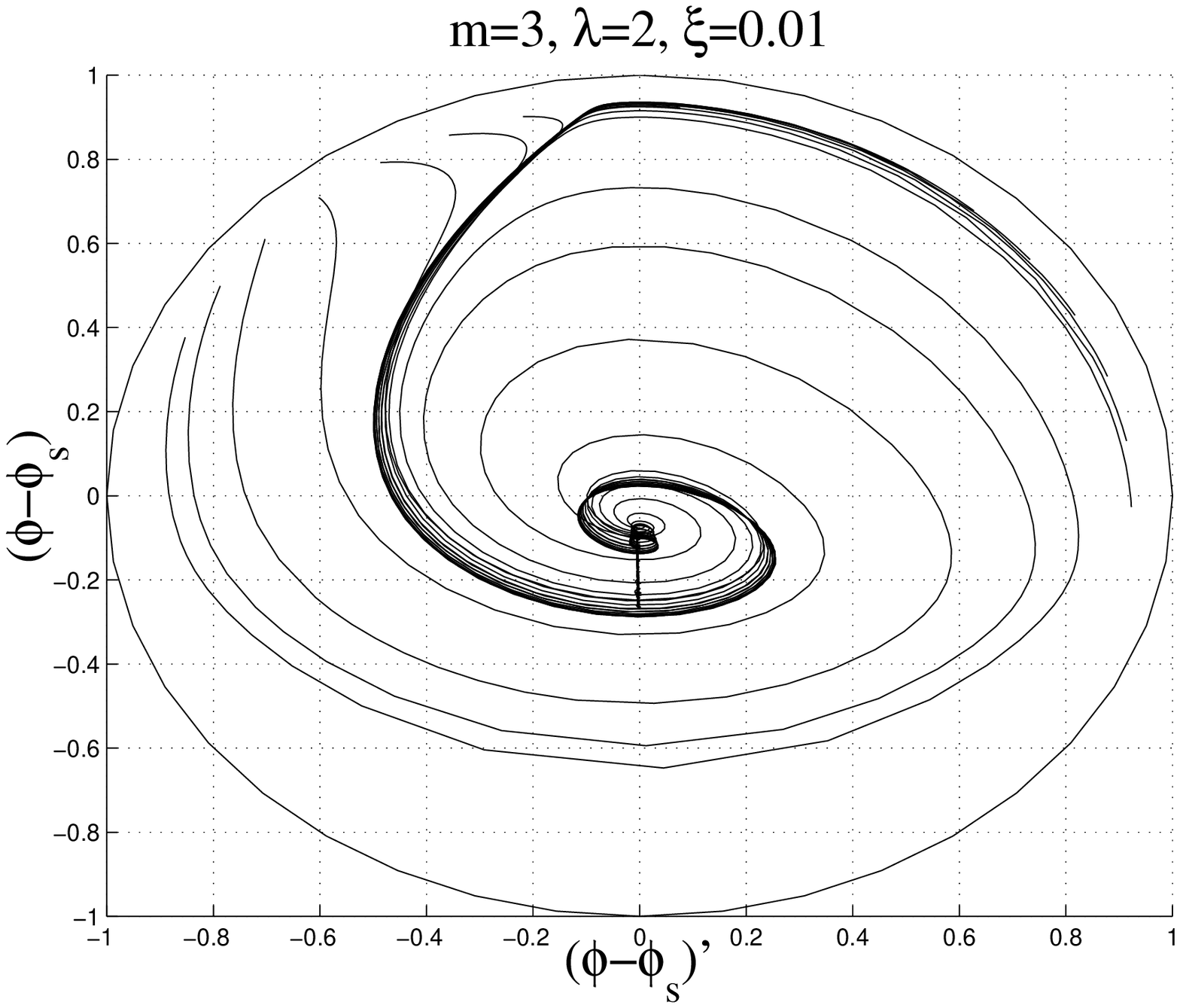, width=5cm}
\epsfig{figure=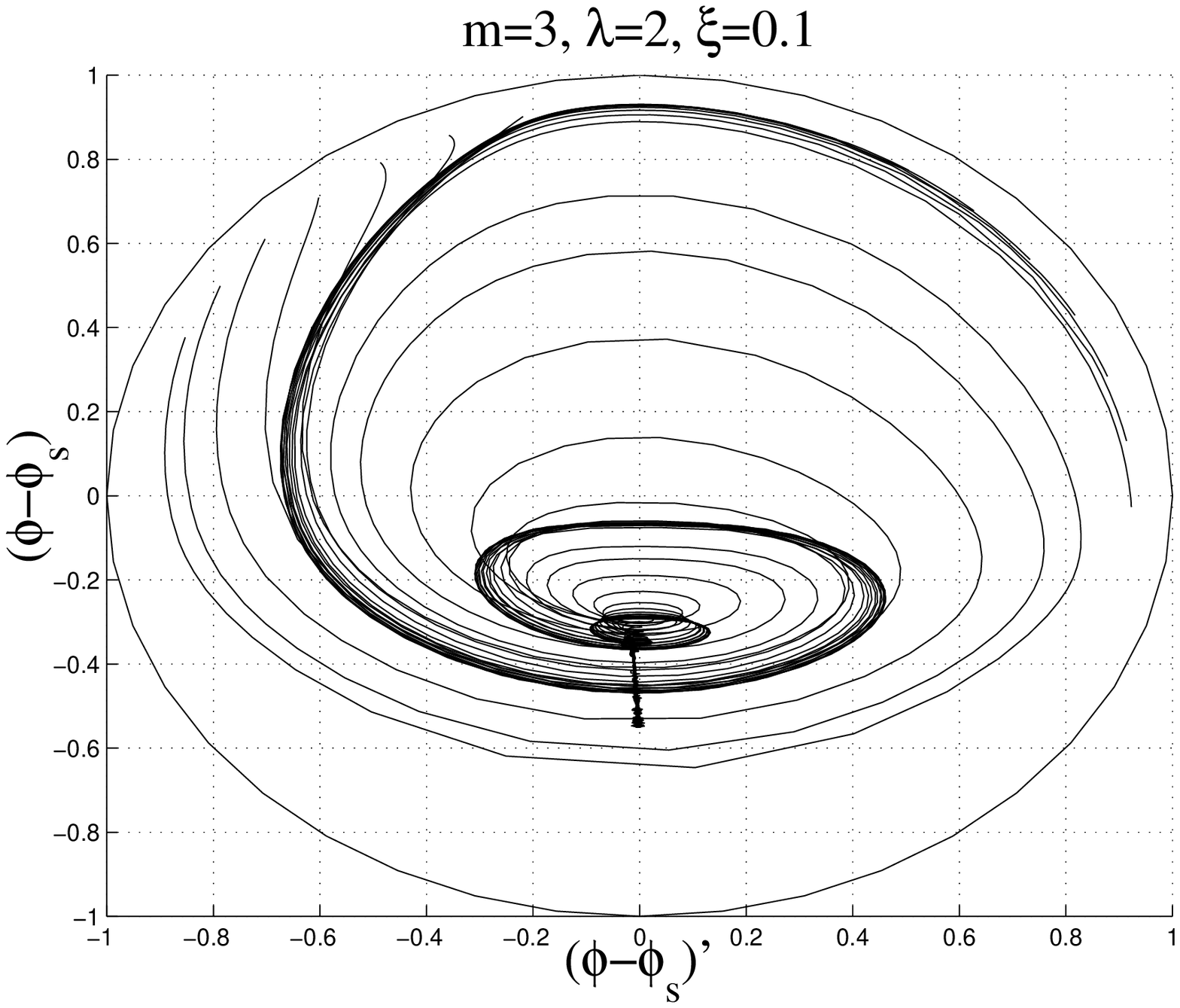, width=5cm}
\epsfig{figure=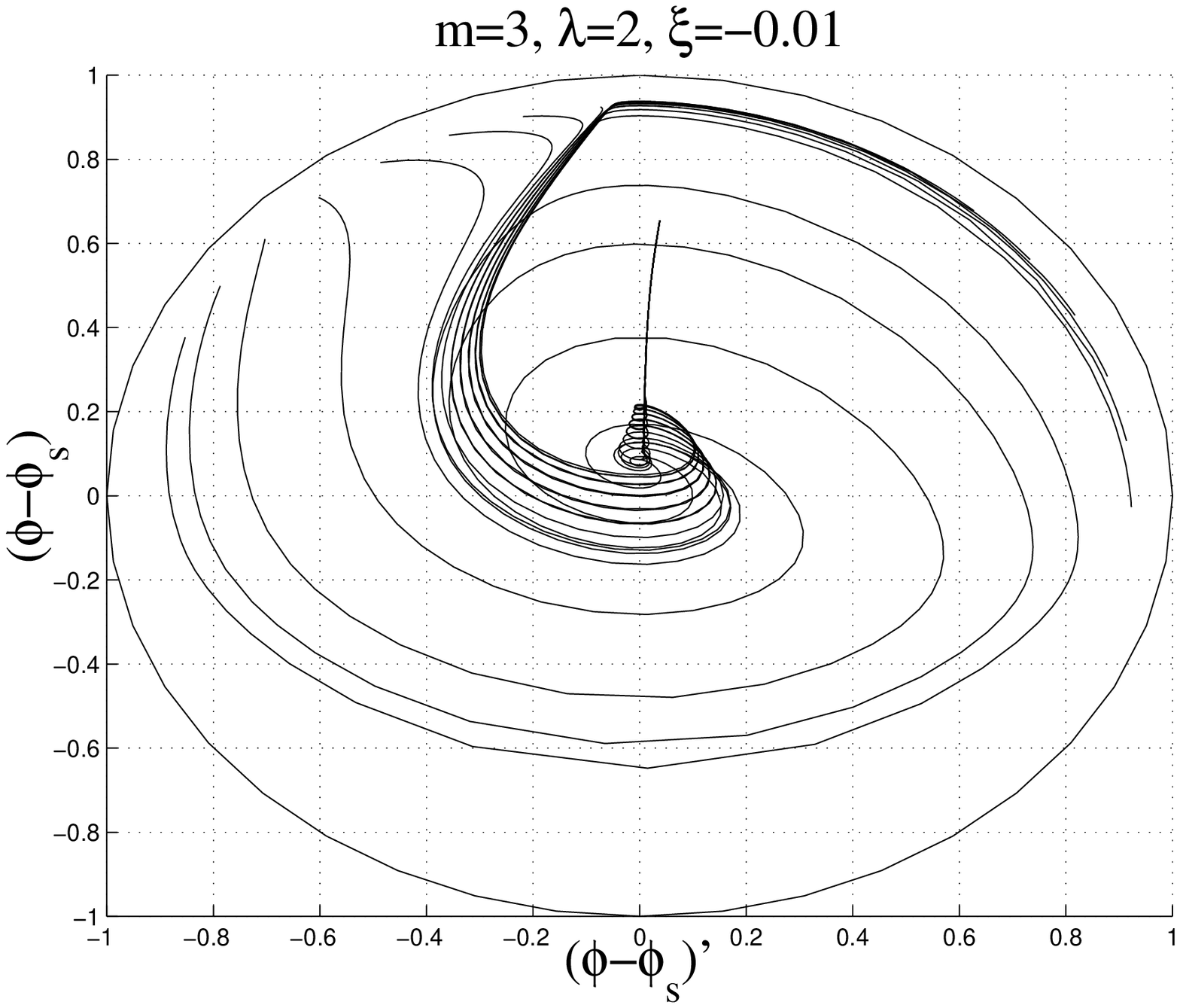, width=5cm}
\caption{The deformation of the phase space from figure \ref{fig1} due
to the coupling for $\xi=10^{-2}, 10^{-1}, -10^{-2}$. The solution
converges towards the attractor which drifts from $\phi_s$, since
$\phi-\phi_s\propto \xi u +{\cal O}(\xi^2)$.}
\label{fig2}
\end{figure}

Indeed, this analysis is valid only as long as the potential
term dominates over the coupling term in equation (\ref{kg_exp}), that is,
as long as
\begin{equation}
\left|\frac{dV}{d\phi}\right|\gg\left|\frac{12}{m}\left(\frac{4}{m}-1\right)
\xi\phi
\hbox{e}^{-2u}\right|\Longleftrightarrow u\ll u_{eq},\quad
u_{eq}\equiv \frac{1}{12|\xi|}\frac{m(6-m)}{|4-m|}\ln{\frac{1}{\lambda}
\sqrt{2\left(\frac{6}{m}-1\right)}}.
\end{equation}
We recover that when $\xi\rightarrow0$ or $m\rightarrow4$,
$u_{eq}\rightarrow\infty$ and we are back to the minimally coupled
case (\S \ref{exp}). The behaviour of $\left|\frac{dV}{d\phi}\right|$
and $\frac{12}{m}\left(\frac{4}{m}-1\right)\xi\phi$ in function of $u$
is shown on figure \ref{fig4}. When $\xi\ll1$, we see that, as
expected, the solution is first dominated by the potential term but
that, as time elapses the coupling term tends to become more
dominant. In figure \ref{fig2}, we illustrate how the phase space
trajectories are deformed due to the existence of this small coupling.

\begin{figure}
\centering
\epsfig{figure=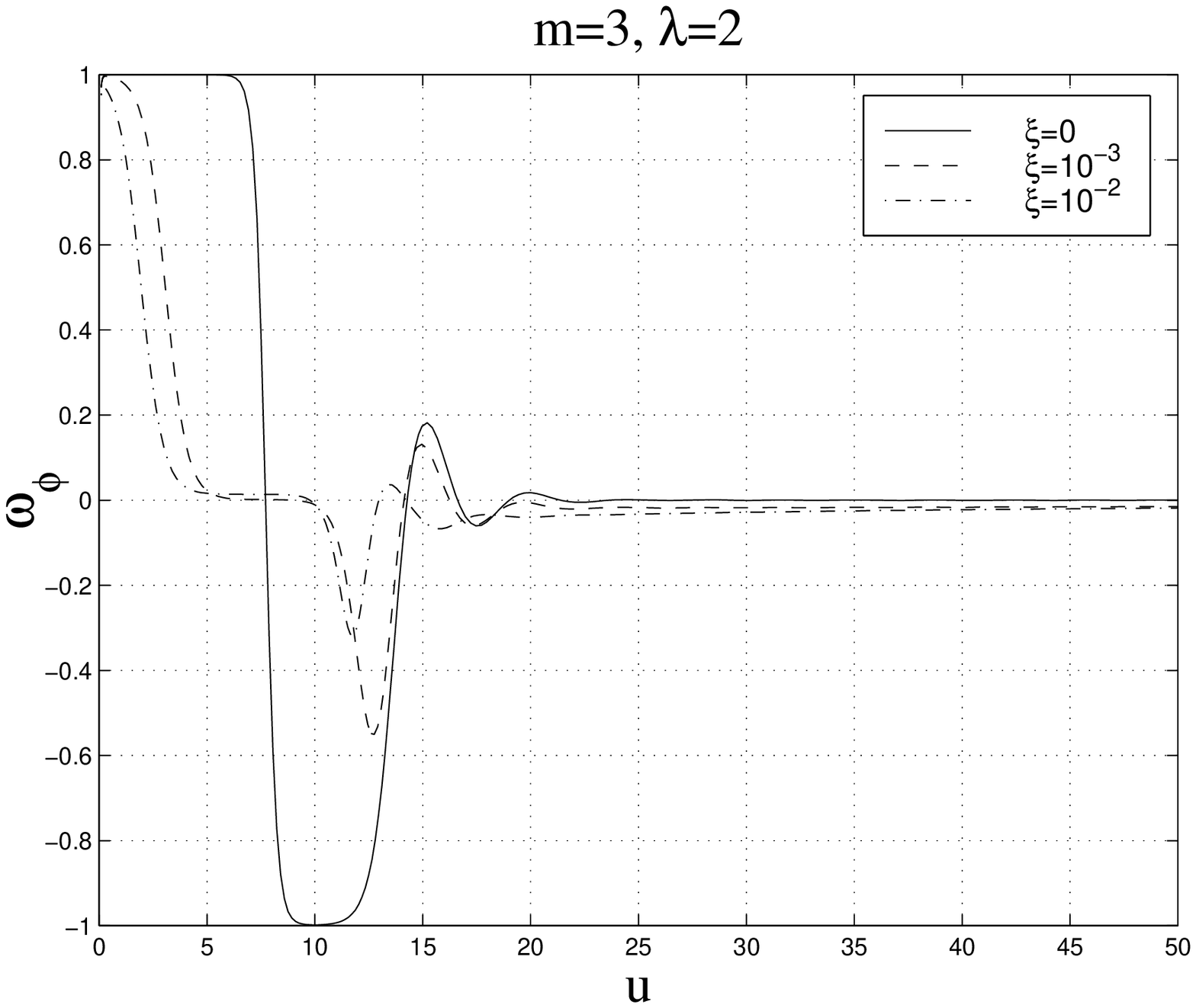, width=6cm}
\epsfig{figure=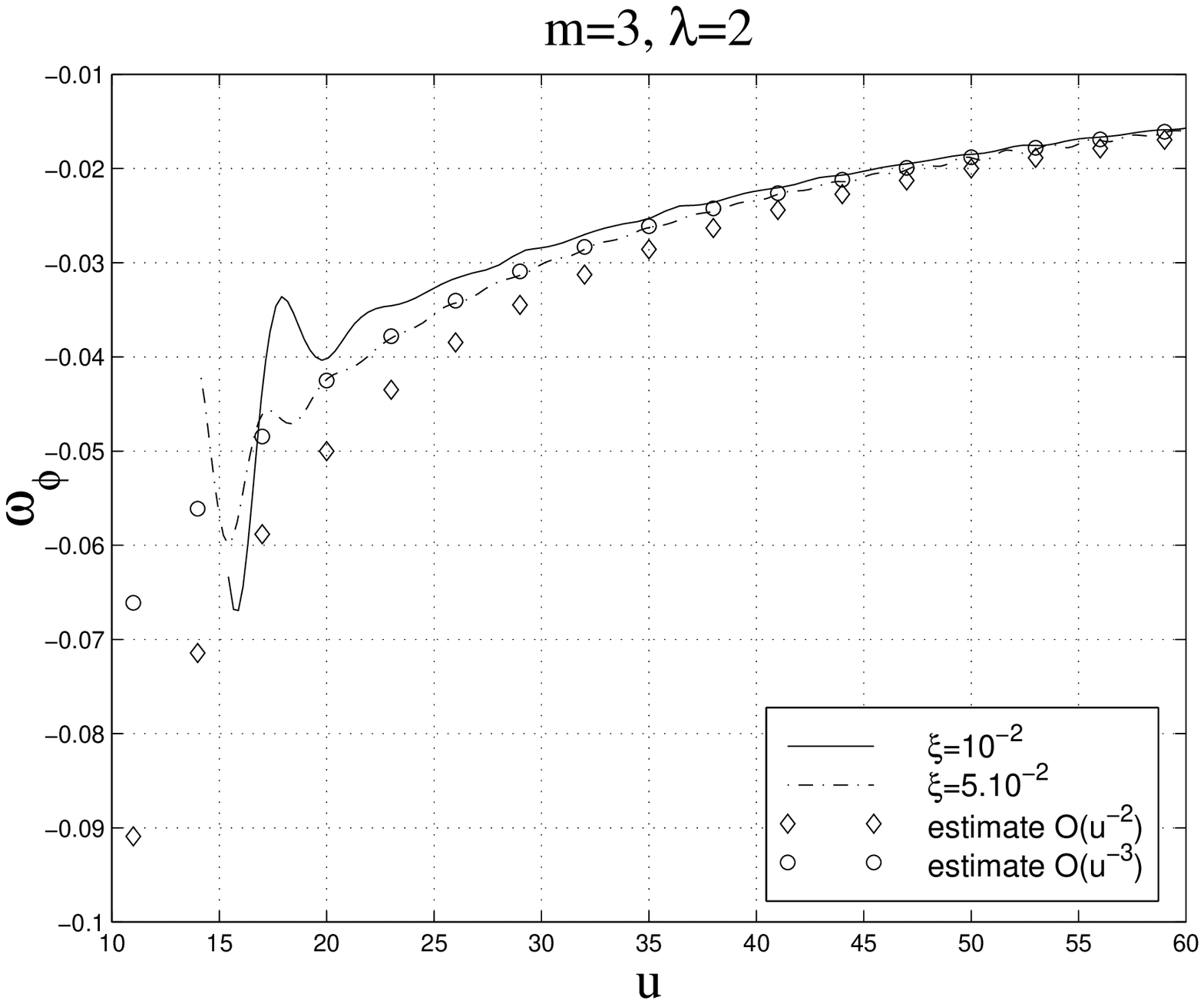, width=6cm}
\caption{The time varying equation of state and its deviation from the
minimally coupling case for $\xi=0$, $10^{-3}$, $10^{-2}$ and the
comparaison of this equation of state with the estimate (\ref{t2}) at
first order and second order in $u^{-1}$.}
\label{fig3}
\end{figure}

\begin{figure}
\centering
\epsfig{figure=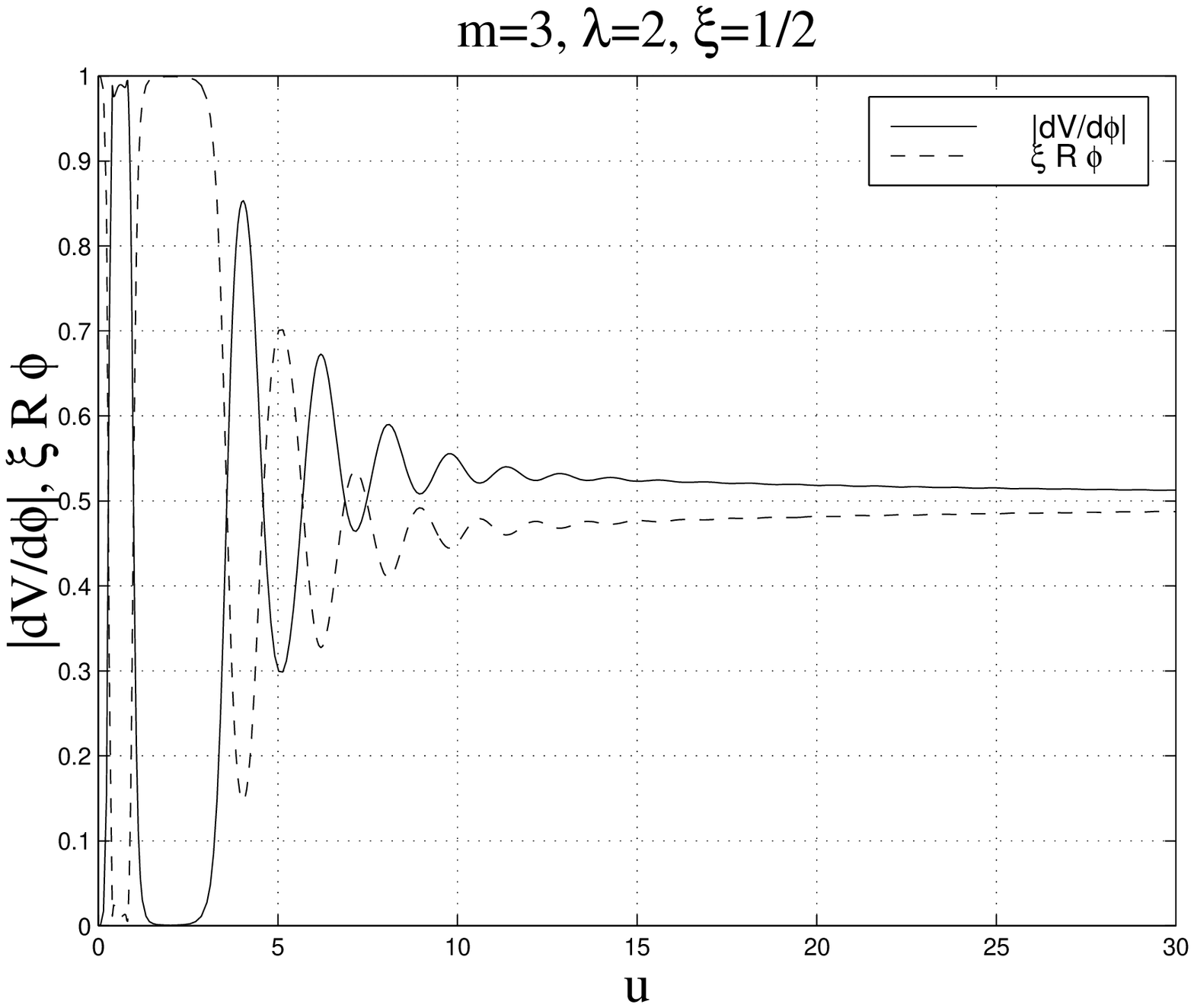, width=5cm}
\epsfig{figure=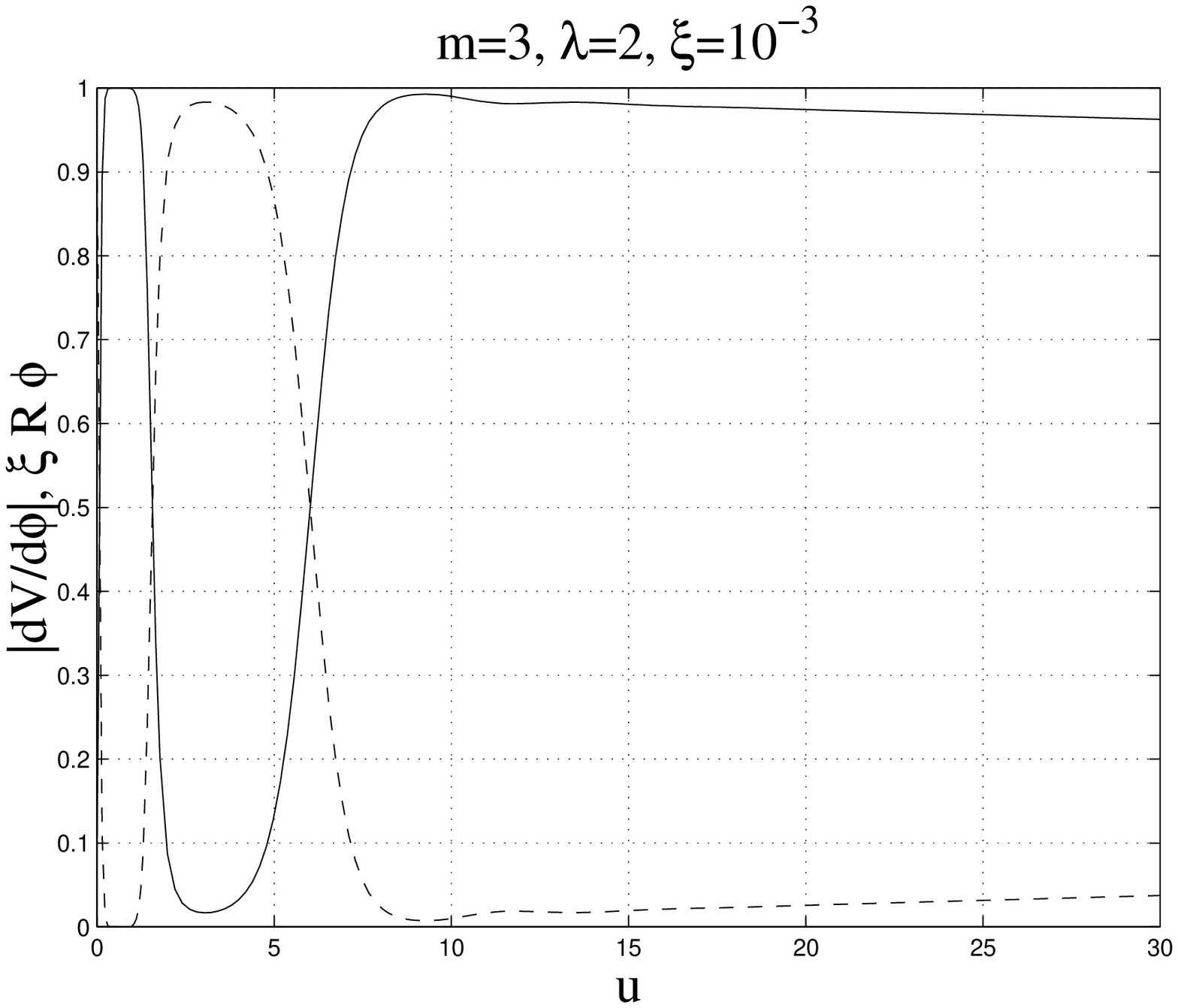, width=5cm}
\epsfig{figure=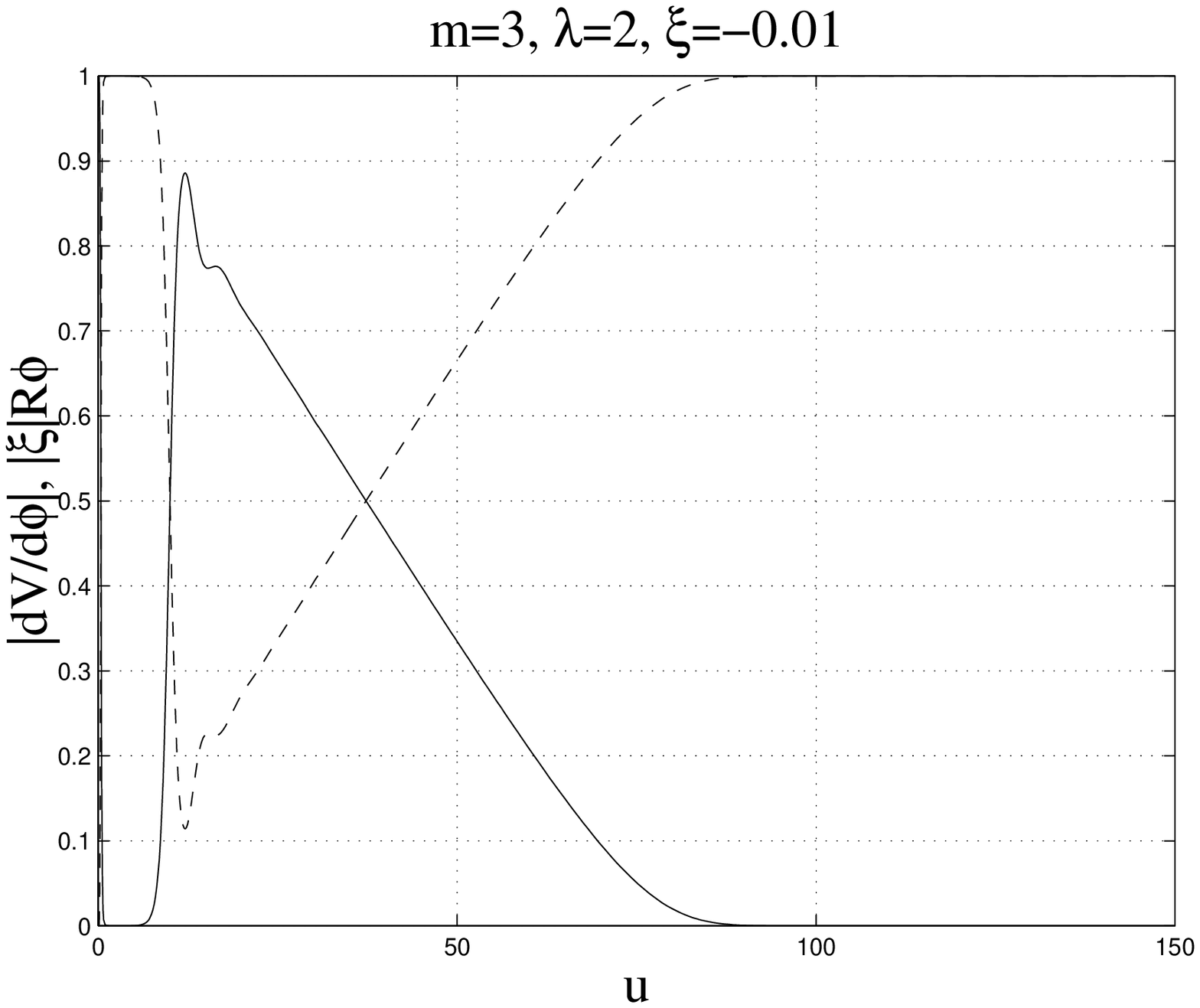, width=5cm}
\caption{The respective influence of the coupling term and the potential term
in the Klein-Gordon equation. We have plotted $\left|\frac{dV}{d\phi}\right|$
and $\left|\bar{\cal R}\xi\phi\right|$ normalised
to their sum for $\xi=1/2$ (left). The two terms alternatively dominate
and then converge. When $\xi\ll1$, after some oscillations
corresponding to the convergence towards $\phi_s$, the potential term
dominates over the coupling as long as $u\ll u_{eq}$. When $\xi<0$,
the coupling term will dominates forever.}
\label{fig4}
\end{figure}

Now, as long as $u\ll u_{eq}$, we can compute the equation of state of
the field by inserting the particular part of the solution
(\ref{toto}) in (\ref{rhophi}-\ref{pphi}) which will take the form
\begin{eqnarray}
\omega_\phi(u)&=&\omega_{\phi_s}\left(1+{\cal
A}(m,\lambda)u^{-1}+{\cal B}(m,\lambda,\xi)u^{-2} +{\cal
O}(u^{-3}\right) \quad\hbox{if}\quad m\not=3\nonumber\\
&=&-\frac{1}{u}\left(1+{\cal C}(m,\lambda)u^{-1}+{\cal O}(u^{-2})\right)
\quad\hbox{if}\quad m=3.\label{t2}
\end{eqnarray}
The difference in these two behaviours comes from the fact that
$P_\phi\propto u^2$ if $m\not=3$ and $P_\phi\propto u$ when $m=3$.
The exact forms of the functions ${\cal A}$, ${\cal B}$ and ${\cal C}$
can be obtained by doing an expansion of $P_\phi/\rho_\phi$ in
$u^{-1}$.  In figure \ref{fig3}, we show the deviation of the equation
of state from pure scaling and compare the former expansion to the
numerical integration. In conclusion we have that when $\xi\ll1$ and
$u\ll u_{eq}$, the field converges towards a barotropic fluid.

\subsubsection{The $|\xi|\gg1$ case}\label{petit1}

Let us now consider the case where initially the coupling term
dominates over the potential term in the Klein-Gordon equation. At
lowest order, one has
\begin{equation}
\phi''+\left(\frac{6}{m}-1\right)\phi'+\frac{12}{m}
\left(\frac{4}{m}-1\right)\xi\phi=0,
\end{equation}
the solution of which is of the form
\begin{equation}
\phi=A_+\hbox{e}^{\alpha_+u}+A_-\hbox{e}^{\alpha_-u}.\label{toto2}
\end{equation}
When $\xi>0$, since $\alpha_+\alpha_-\propto4/m-1$ and $\alpha_++\alpha_-=-(6/m-1)$,
we deduce that if $0<m<4$, the real part of both $\alpha_+$ and
$\alpha_-$ are negative so that (\ref{toto2}) corresponds to two
decaying modes [It can be seen as a proof that the critical
point $\phi=0$ is an attractor] so that
\begin{equation}
\bar{\cal R}\xi\phi\rightarrow0\quad\hbox{and}\quad
\lambda\hbox{e}^{-\lambda\phi}\rightarrow\lambda\quad\hbox{when}\quad0<m<4
\quad\hbox{and}\quad \xi>0,
\end{equation}
and the potential term will rapidly catch up the coupling term.

When $\xi<0$, $\alpha_+\alpha_-\propto-4/m+1$ and the real part of one of the
two quantities, $\alpha_+$ say, is positive when $0<m<4$ so that
\begin{equation}
\bar{\cal R}\xi\phi\rightarrow\infty\quad\hbox{and}\quad
\lambda\hbox{e}^{-\lambda\phi}\rightarrow0\quad\hbox{when}\quad0<m<4
\quad\hbox{and}\quad \xi<0,
\end{equation}
and the coupling term will dominates forever (see figure \ref{fig4}) and
$\phi$ will behave as $A_+\hbox{e}^{\alpha_+u}$ with
\begin{equation}
2\alpha_+=\left(1-\frac{6}{m}\right)+\sqrt{
\left(\frac{6}{m}-1\right)^2-\frac{48}{m}\left(\frac{4}{m}-1\right)\xi},
\label{rac}
\end{equation}
for which, since the potential term is negligible,
\begin{equation}
\omega_\phi(m,\xi)\simeq\frac{\alpha_+^2-4\xi\left[\frac{2}{m^2}(3-m)+
\left(\frac{4}{m}-1\right)\alpha_++2\alpha_+^2\right]}
{\alpha_+^2+\frac{24}{m}\xi\left[\alpha_++\frac{1}{m}\right]}.
\end{equation}
This solution has however to be excluded since one
can check that it leads to $\rho_\phi<0$.

\subsubsection{General case}

The general case is more involved since we cannot find any analytic
solution to (\ref{kg_exp}). First, when $\xi>0$, we have seen that
when either the potential or the coupling term dominates, the other
slowly catch up. We thus expect a late time solution which satisfies
\begin{equation}
\bar{\cal R}\xi\phi\simeq\lambda\hbox{e}^{-\lambda\phi}.
\end{equation}
In figure \ref{fig4}, we plot the evolution of these
two terms in the case where $\xi\ll1$ and in a more
general case. The two terms alternatively dominate and then converge
to the same value. Indeed, this is no proof.\\

We can however look for a general solution of the form
\begin{equation}
\phi=\sum_{n=0}^{n=\infty} \xi^n\psi_n,
\end{equation}
with $\psi_0$ given by (\ref{sol_exp}) and $\psi_1$ given by (\ref{toto}).
Inserting this expansion in the Klein-Gordon equation (\ref{kg_exp}),
we obtain the hierarchy of equations
\begin{equation}
\psi''_n+\left(\frac{6}{m}-1\right)\psi'_n+2\left(\frac{6}{m}-1\right)\psi_n
= -\frac{12}{m}\left(\frac{4}{m}-1\right)\psi_{n-1} +\lambda A^{-\lambda}
f_n(\psi_0,...,\psi_{n-1})\quad\hbox{if}\quad n\geq1.
\end{equation}
The functions $f_n$ depends on all the solutions $\psi_i$ for
$i<n$. All these equations have a solution of the form
\begin{equation}
\psi=B_+\hbox{e}^{\alpha_+u}+B_-\hbox{e}^{\alpha_-u}+\hat\psi_n.
\end{equation} 
As in \S \ref{petit1}, the two homogeneous modes decay if $m<6$ and
$\forall n, \psi_n\rightarrow\hat\psi_n$. It can also be seen that
$\hat\psi_n$ will be a polynomial in $u$ of degre $n$. If the series
$\sum\xi^n\hat\psi_n$ converges then,
\begin{equation}
\phi\rightarrow\sum_{n=0}^{n=\infty}\xi^n\hat\psi_n\quad\hbox{when}
\quad u\rightarrow\infty,\label{tt2}
\end{equation}
from which we can conclude that if $\sum\xi^n\hat\psi_n$ converges,
there exists an attractor to the equation (\ref{kg_exp}) given
by (\ref{tt2}). Indeed, we cannot demonstrate the convergence
of this series in the general case. In figure \ref{fig6}, we show
the phase space trajectories showing the convergence towards this attractor.

When $m>2/3$, the solution converges towards the attractor while oscillating
so that the equation of state have wiggles and converges towards $\omega_{\phi_s}$
(see figure \ref{fig6}). When $m<2/3$, this oscillatory behaviour
does not appear whatever $\xi$ (see figure \ref{fig6}).\\

\subsubsection{Case of a radiation dominated universe}

\begin{figure}
\centering
\epsfig{figure=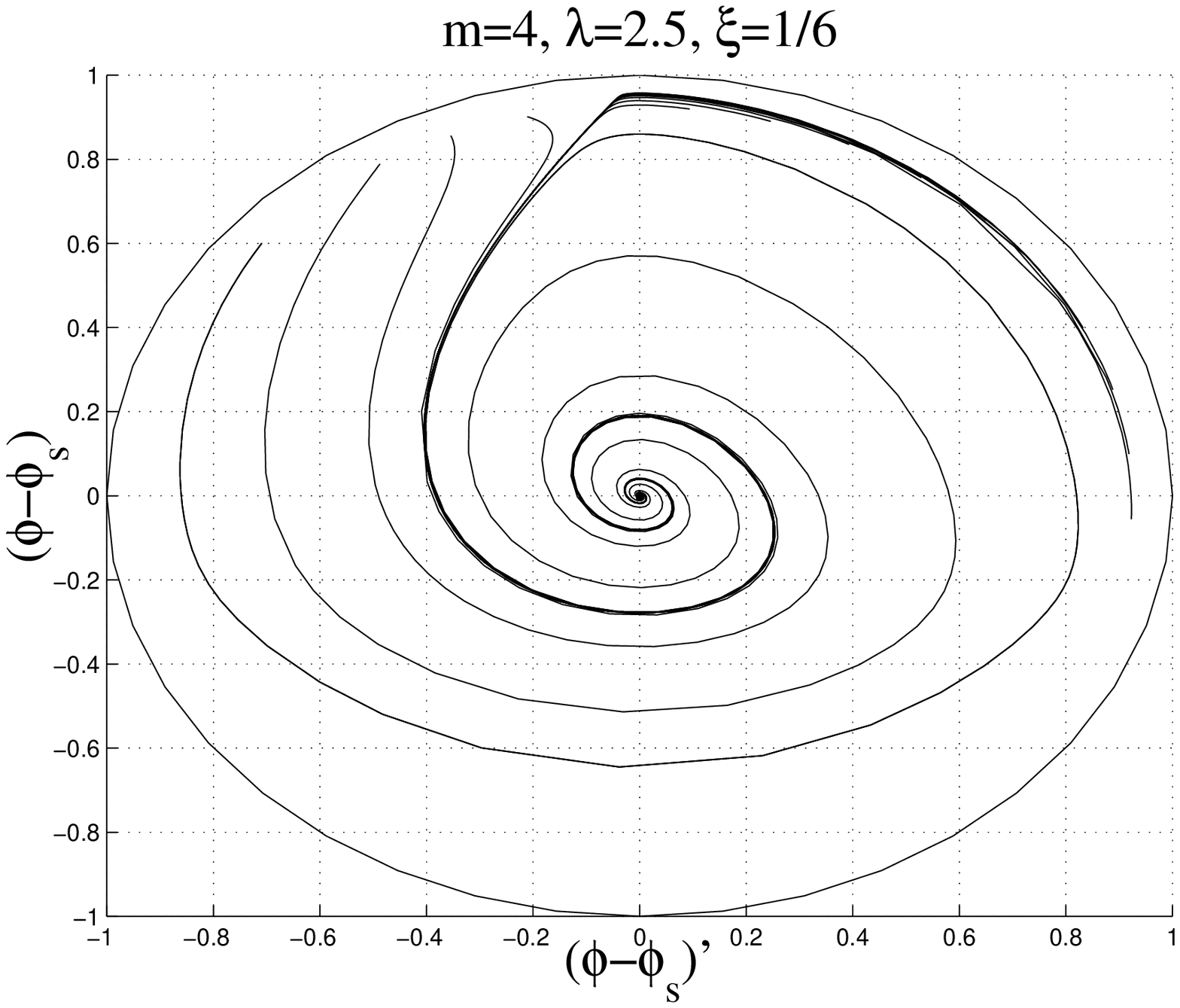, width=6cm}
\epsfig{figure=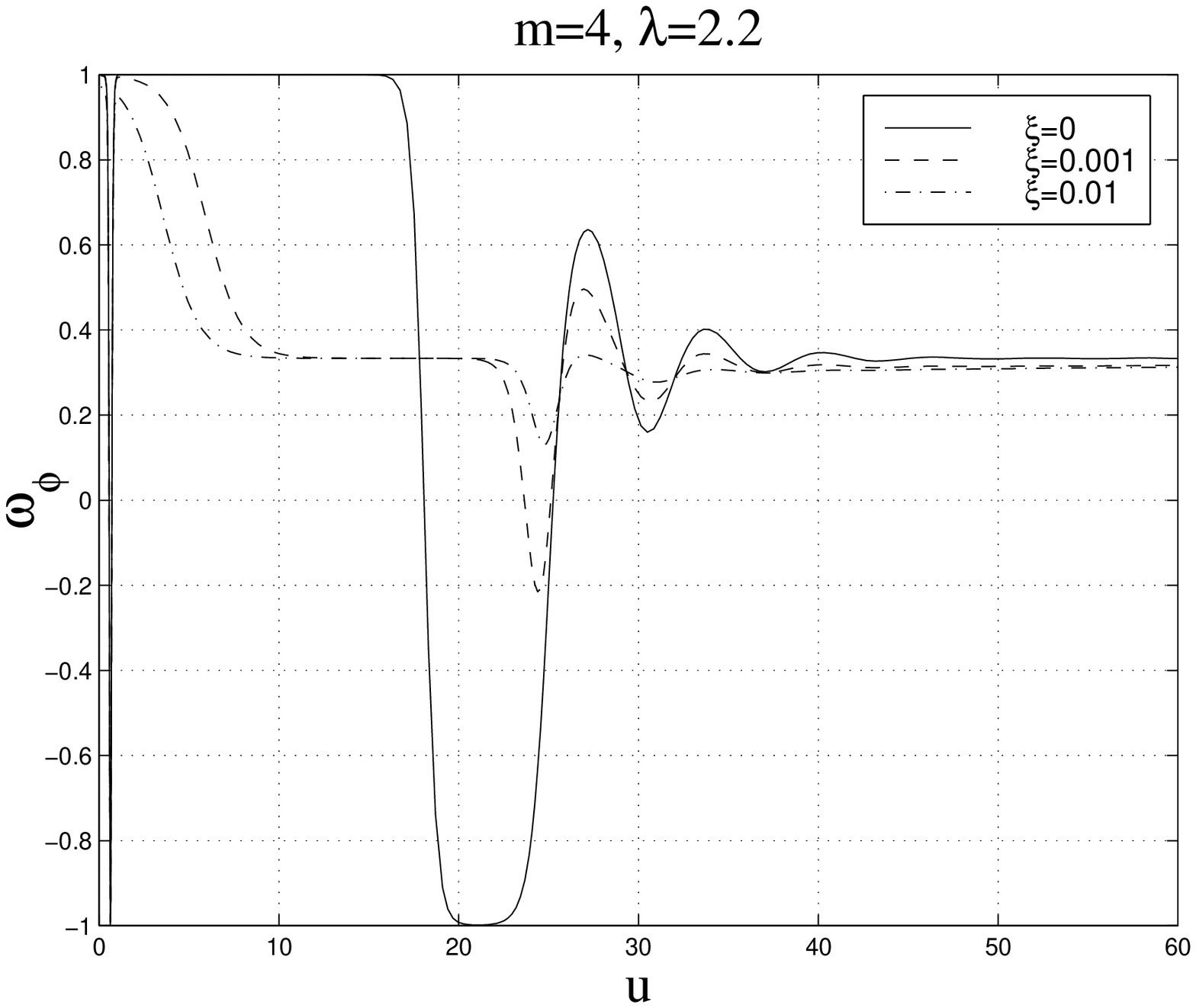, width=6cm}
\caption{The phase space analysis for a field in a radiation dominated universe.
The scaling solution is the late time attractor whatever $\xi$
[$\xi=0$, $10^{-2}$ and  $10^{-3}$].}
\label{fig5}
\end{figure}

\begin{figure}
\centering
\epsfig{figure=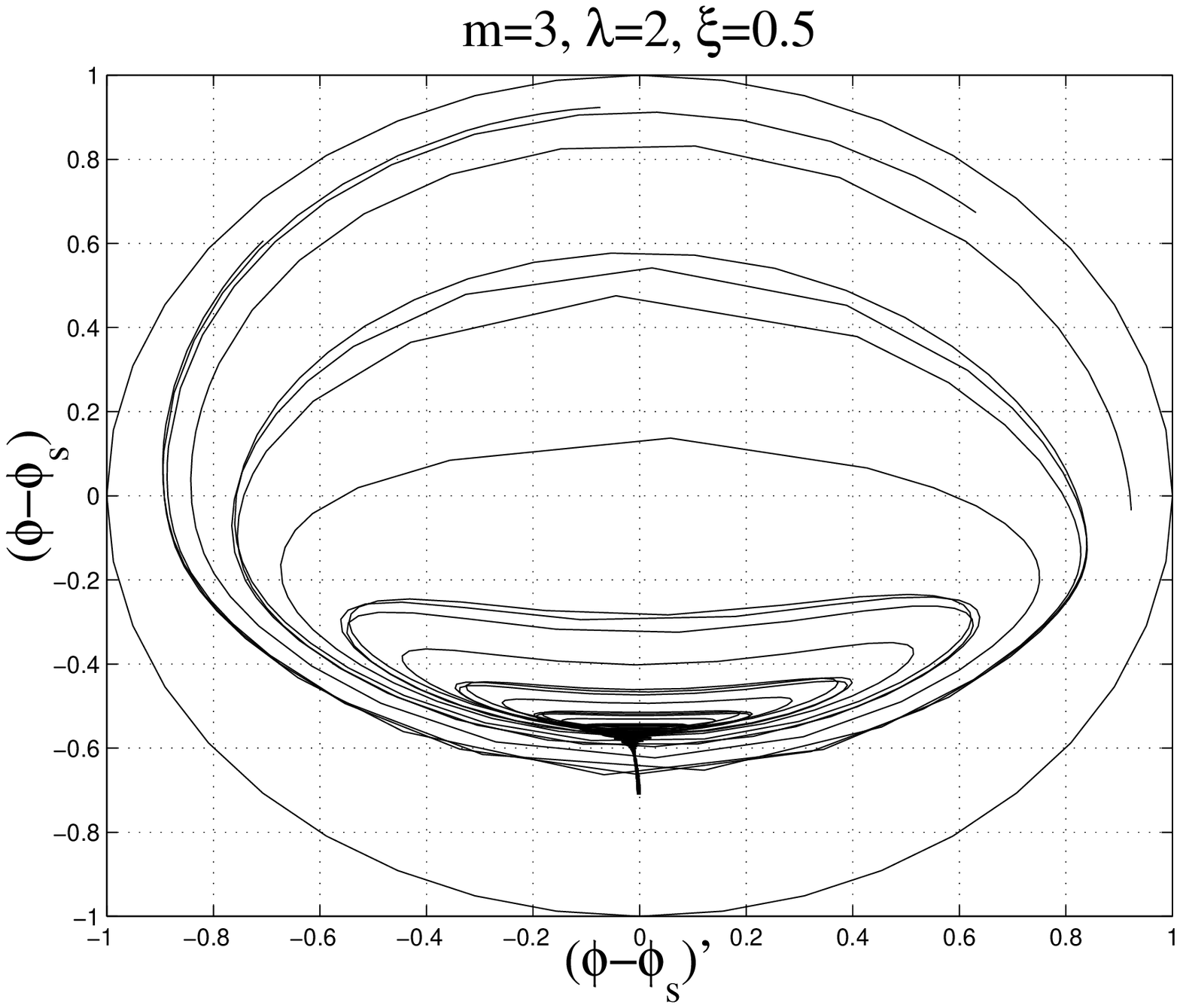, width=6cm}
\epsfig{figure=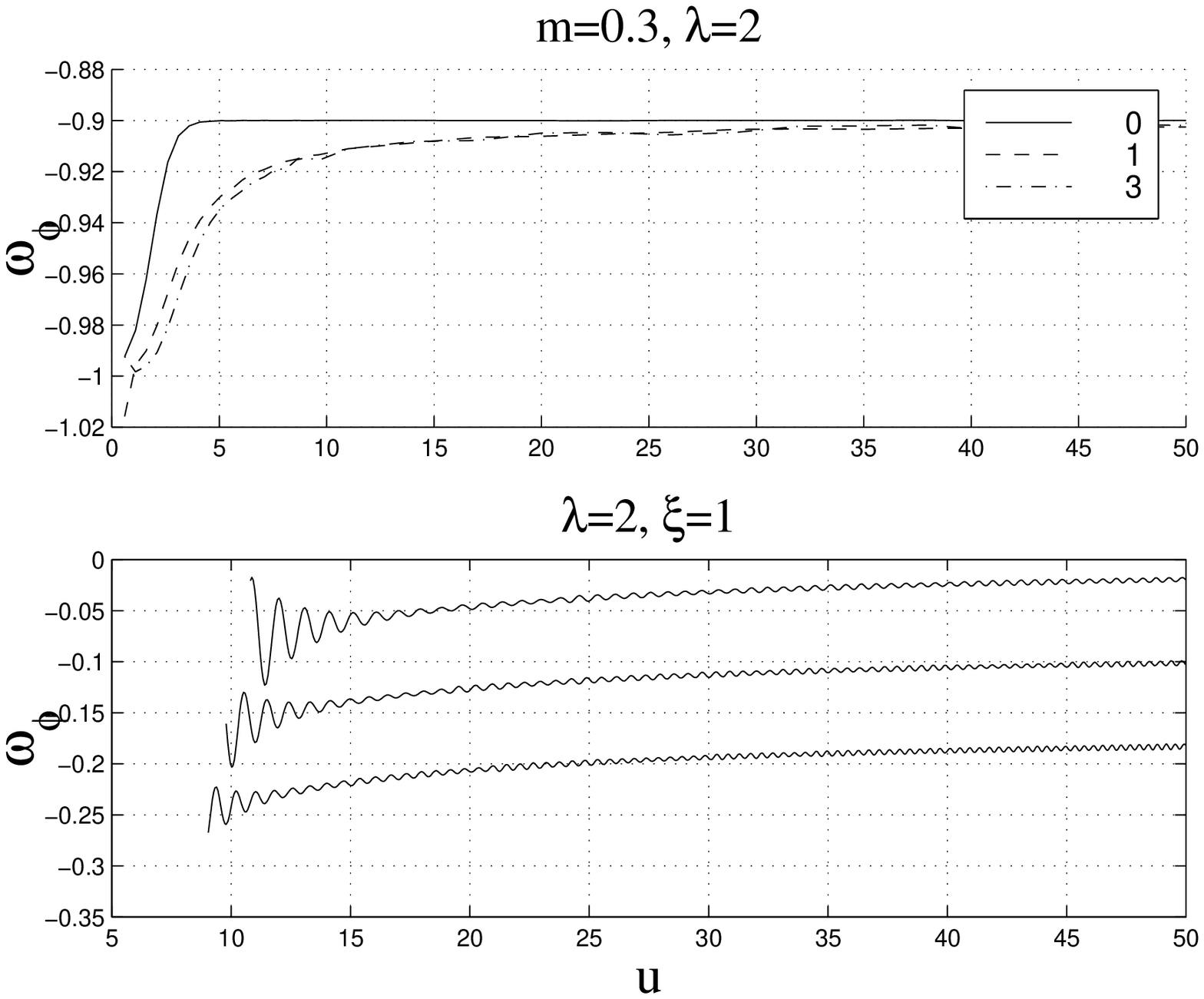, width=6cm}
\caption{The phase space analysis and the evolution of the density, pressure and
equation of state in a case where $\xi\not\ll1$. $\omega_\phi$ 
converges towards $\omega_{\phi_s}$ whatever the value of $\xi$
with or without oscillations according to the value of $m$ (m=2, 2.5, 3
from bottom to top in the lower right plot).}
\label{fig6}
\end{figure}

As explained above, one has the solution
(\ref{sol_exp}) to the Klein-Gordon equation and this solution is an
attractor (see the phase analysis of figure \ref{fig5}) and the field
behaves as radiation. The solution (\ref{toto}) reduces to the two
decaying modes. Indeed the derivation of (\ref{t2}) is no longer valid
but, with the same method, one can show that, to first order in
$u^{-1}$,
\begin{equation}
\omega_\phi=\frac{1}{3}\left(1+{\cal A}(\lambda)u^{-1}+{\cal
B}(\lambda,\xi)u^{-2} \right),
\end{equation}
where the two functions ${\cal A}$ and ${\cal B}$ are obained as in \S
\ref{petit1}. Note that for $u\gg1$ the $\xi$ terms of
(\ref{rhophi}-\ref{pphi}) dominates the density and the pressure. In
figure \ref{fig5}, we have plotted the phase space showing the
convergence towards the attractor $\phi_s$ in the case of a conformally
coupled scalar field and the evolution of its equation of state.

\subsection{Numerical results}

We integrate numerically equation (\ref{kg_exp}) and use the Poincar\'e
projection \cite{ellis} to represent the result in the plane $(\phi,\phi')$
\begin{eqnarray}
\phi&=&\frac{r}{1-r}\sin{\theta}\nonumber\\
\phi'&=&\frac{r}{1-r}\cos{\theta}.
\end{eqnarray}
This projection shrinks all the trajectory in the phase-space to the
unit disk. The point $N\equiv(0,1)$, $E\equiv(-1,0)$ and
$W\equiv(1,0)$ respectively correspond to ($\phi=\infty,\phi'=0$),
($\phi=0,\phi'=-\infty$) and ($\phi=0,\phi'=\infty$). Note that the
system is non-autonomous since equation (\ref{exp_int}) depends on
$u$. Thus, curves can cross in the Poincar\'e representation but they
will not cross in a 3D representation $(\phi,\phi',u)$. The following
plots corresponds to the situation studied in the previous paragraphs.
We plot the evolution of the solution in the phase space $(\phi,\phi')$ and
the evolution of the equation of state when $\xi\ll1$. To finish,
we give some examples of evolution in the case $\xi\not\ll1$.

\section{Conclusion}

In this article, we studied the influence of the coupling between the
scalar curvature and the scalar field on the existence and stability
of scaling solution of this field evolving in either an exponential or
an inverse power law potential. The motivation for considering such
solutions is first that they can be a candidate for a matter
component with negative pressure and second that they appear
for a large class of potentials predicted by some theories
of hight energy physics.

We first found a new parametric form of the potential that reduces 
to the inverse power law potential when the energy density
of the fluid drives the evolution of the spacetime.

Concerning the inverse power law potentials, we showed analytically
that the existence and stability of a scaling solution does not depend
on the coupling $\xi$ and that the equation of state of the field was
always given by $\omega_\phi=\frac{\omega_{_B}\alpha-2}{\alpha+2}$. This
generalises the work by Ratra-Peebles \cite{ratra88} and
Liddle-Scherrer \cite{liddle98}.

The situation is more involved with exponential potentials since one
cannot find an analytic form for a scaling solution (apart for a
radiation dominated universe). We then studied the effect of a small
perturbation coupled to the scalar curvature and computed
$\omega_\phi$ to first order in $\xi\ll1$ and showed that there always
exists a time after which one cannot neglect the effect of the
coupling. In that limit, we show that the equation of state was
converging towards a barotropic equation of state and some numerical
examples tend to show us that it should be the case whatever $\xi>0$
(but this has not been demonstrated).  Indeed, such potentials are not
the most favored since they are constrained by nucleosynthesis to
$\Omega_{\phi0}<0.15$ \cite{ferreira97} and cannot explain (when
$\xi=0$) the supernovae measurements since $\omega_\phi=0$. Note
however that the convergence toward the barotropic equation
$\omega_{\phi_s}$ is much slower when $\xi\not=0$.  When $\xi<0$, we
have shown that there always exists a time after which the 
coupling term will dominate and thus that there exists a scaling
solution but with a different equation of state (as long as $0<m<4$,
and thus in a matter dominated era). Unfortunately, such a solution
has to be rejected since it has negative energy.

As pointed out by Caldwell {\em et al.} \cite{caldwell97}, a smooth
time dependent field is unphysical since ``one has to take into account
the back-reaction of the fluctuations in the matter components''. The
cosmological consequences of such an inhomogeneous coupled scalar
field in both exponential and inverse power law potentials (such as
the computation of the CMB anisotropies and the matter power spectrum)
will be presented later \cite{prepa} and our present study is
only related to the implication of the homogeneous part of such a 
field.

\section*{Aknolwdegments}

It is a pleasure to thank Pierre Bin\'etruy who raised my interest to
the subject of quintessence, Luca Amendola, Nuno Antunes, Nathalie
Deruelle and Alessandro Melchiorri for discussions, and
Ruth Durrer and Patrick Peter for discussions and comments
on the early versions of this text.



\end{document}